\documentclass[aps,twocolumn,prl,showpacs,nofootinbib]{revtex4}

\usepackage{amsmath}
\usepackage{graphicx}
\usepackage{subfigure}
\usepackage{dcolumn}
\usepackage{bm}
\usepackage{amssymb}
\usepackage{latexsym}

\newcommand{\ie}{\textsl{i.e.~}}

\def\spose#1{\hbox to 0pt{#1\hss}}
\def\lta{\mathrel{\spose{\lower 3pt\hbox{$\mathchar"218$}}
     \raise 2.0pt\hbox{$\mathchar"13C$}}}
\def\gta{\mathrel{\spose{\lower 3pt\hbox{$\mathchar"218$}}
     \raise 2.0pt\hbox{$\mathchar"13E$}}}

\newcommand{\de}[2]{\kern - #1 em \mathrm{d} #2}

\begin{document}

\hskip 15.2cm \mbox{\tt RESCEU-10/09}

\title{The moduli problem at the perturbative level}

\author{Martin Lemoine} \email{lemoine@iap.fr} \affiliation{Institut
d'Astrophysique de Paris, CNRS - UPMC, 98bis boulevard Arago, 75014 Paris, France}

\author{J\'er\^ome Martin} \email{jmartin@iap.fr}
\affiliation{Institut
d'Astrophysique de Paris, CNRS - UPMC, 98bis boulevard Arago, 75014 Paris, France}

\author{Jun'ichi Yokoyama} \email{yokoyama@resceu.s.u-tokyo.ac.jp}
\affiliation{Research Center for the Early Universe (RESCEU), Graduate
  School of Science, The University of Tokyo, Tokyo, 113--0033,
  Japan}
\affiliation{Institute for the Physics and Mathematics of the Universe (IPMU)
The University of Tokyo, Kashiwa, Chiba 277-8568, Japan}

\date{\today}

\begin{abstract}
  Moduli fields generically produce strong dark matter -- radiation
  and baryon -- radiation isocurvature perturbations through their
  decay if they remain light during inflation. We show that existing
  upper bounds on the magnitude of such fluctuations can thus be
  translated into stringent constraints on the moduli parameter space
  $m_\sigma$ (modulus mass) -- $\sigma_{\rm inf}$ (modulus vacuum
  expectation value at the end of inflation). These constraints are
  complementary to previously existing bounds so that the moduli
  problem becomes worse at the perturbative level. In particular, if
  the inflationary scale $H_{\rm inf}\sim 10^{13}\,$GeV, particle
  physics scenarios which predict high moduli masses
  $m_\sigma\,\gtrsim\,10-100\,$TeV are plagued by the perturbative
  moduli problem, even though they evade big-bang nucleosynthesis
  constraints.

\end{abstract}


\maketitle

{\it Introduction.}  The phenomenology of moduli fields of
supersymmetric and superstring theories, or more generally, of late
decaying flat directions of the scalar potential is strongly
constrained by cosmological
arguments~\cite{Coughlan:1983ci,Goncharov:1984qm}. Most notably,
successful big-bang nucleosynthesis (BBN) sets very stringent upper
bounds on their energy density at temperatures
$T\,\lesssim\,1\,$MeV~\cite{Kawasaki:2004qu}. Such a small modulus
abundance can be achieved through the dilution of the pre-existing
moduli by late time entropy production or late
inflation~\cite{Randall:1994fr,Lyth:1995ka}; or, if the supergravity
induced potential is such that the moduli has always stayed close to
its low energy minimum~\cite{Dine:1995uk,Dine:2000ds}.  Alternatively,
the BBN constraints can be evaded if the moduli are massive enough to
decay at high temperatures $T_{\rm d}\,\simeq\, 2.8\,{\rm
  MeV}\,(m_\sigma/100\,{\rm TeV})^{3/2}$, a property which is
considered as a strong motivation for particle physics models which
realize this hierarchy $m_\sigma\,\gtrsim\,10-100\,{\rm
  TeV}\,\gg\,M_W$ (weak scale), such as noscale supergravity, anomaly
mediation or string inspired
models~\cite{Ellis:1986zt,Kawasaki:1995cy,Moroi:1999zb,Acharya:2008bk}.

\par

These constraints directly relate to the background evolution of the
modulus field. However, observational cosmology now offers powerful
constraints on the power spectrum of density perturbations and the
nature of these perturbations. As we demonstrate for the first time in
this Letter, these constraints can be translated into stringent
constraints on moduli cosmology which directly impact on high energy
physics model building in the modulus sector. In particular, we find
that the moduli problem becomes worse at the perturbative level,
because the modulus generically produces strong isocurvature
fluctuations between dark matter and radiation and between baryons and
radiation (as noted in~\cite{Moroi:2001ct}), in conflict with existing
cosmic microwave background (CMB) anisotropy data.

\par

The constraints obtained depend on the shape of the effective
potential of the moduli during and after inflation, and can be
formulated most clearly in the parameter space $(\sigma_{\rm
  inf},\,\,m_\sigma)$. In the following we discuss three prototypical
cases in which the modulus mass receives supergravity corrections of
order $H$ or not, and whether the inflaton produces moduli during its
decay or not. We focus here on $m^2\phi^2$ inflation with $H_{\rm
  inf}=10^{13}\,$GeV, as such models provide the most simple
realization of inflation with a scalar index in accord with CMB
measurements~\cite{Boyle:2005ug}. An accompanying paper~\cite{LMY09}
provides the details of the calculations and extends the survey of the
constraints to other possible cases and concrete inflationary model
building.

\par

{\it Isocurvature fluctuations from moduli fields}. Moduli acquire
their own spectrum of density fluctuations through inflation, with
$\delta\sigma_{\rm inf}/\sigma_{\rm inf}\,\sim\,H_{\rm
  inf}/(2\pi\sigma_{\rm inf})$, assuming
that $m_\sigma\,\ll\,H_{\rm inf}$ during inflation. One crucial
observation is that these fluctuations are independent of those of the
radiation, dark matter and baryon fluids, which inherit those of the
inflaton at reheating, with an energy density contrast of order
$10^{-5}$. As the modulus decays, it transfers most of its energy to
radiation, a fraction $B_\chi\,\ll\,1$ to dark matter, and another
part into baryons and anti-baryons. As the final fractional density
perturbations of radiation ($\Delta_\gamma$), dark matter
($\Delta_\chi$) and baryons ($\Delta_{\rm b}$) comprise different
mixtures of the initial inflaton and modulus perturbations,
isocurvature modes between those fluids exist. To go one step further,
a significant dark matter -- radiation isocurvature perturbation is
generated when a significant part of dark matter has been produced by
the modulus while the radiation fluid has remained unaffected, \ie
$B_\chi\,\Omega_\sigma^{<_{\rm d}}\,\gg\,\Omega_\chi^{<_{\rm d}}$ and
$\Omega_\sigma^{<_{\rm d}}\,\ll\,1$ [the superscripts $<_{\rm d}$
(resp. $>_{\rm d}$) mean immediately before (resp. after) modulus
decay, the subscripts $\chi$ (resp. $\sigma $) refer to dark matter
(resp. modulus)]. However, if modulus decay preserves baryon number
(as we assume here), its decay cannot affect the perturbations of
``net baryon number'', hence the baryon isocurvature mode is generated
when the modulus significantly reheats the Universe through its decay,
\ie $\Omega_\sigma^{<_{\rm d}}\not \ll 1$. The fact that dark matter
and baryon isocurvature modes are produced at very different values of
$\Omega_\sigma^{<_{\rm d}}$ explains the power of the constraints
derived in this work.

\par

The final isocurvature fluctuations, to be tested against CMB data are
explicitly written as:
\begin{eqnarray}
S_{\chi\gamma}^{>_{\rm d}}&\,\simeq\,&\frac{1}{1+\Upsilon_\chi}\left(
\frac{B_\chi\Omega_{\sigma}^{<_{\rm d}}}{ \Omega_{\chi}^{<_{\rm d}} +
  B_\chi\Omega_{\sigma}^{<_{\rm d}}}-\Omega_{\sigma}^{<_{\rm d}}\right)
S_{\sigma\gamma}^{\rm (i)}\ ,\label{eq:Scr}\\ 
S_{\rm b\gamma}^{>_{\rm d}}&\,\simeq\,& -\Omega_{\sigma}^{<_{\rm d}}
S_{\sigma \gamma}^{\rm (i)}\ ,\label{eq:Sbr}
\end{eqnarray}
with $S_{\alpha\beta}\,\equiv\,3(\zeta_\alpha-\zeta_\beta)$ (the
subscript $\gamma $ refers to radiation) and $S_{\sigma\gamma}^{\rm
  (i)}\,=\,2\delta\sigma_{\rm inf}/\sigma_{\rm
  inf}$~\cite{Langlois:2004nn}; $\Upsilon_\chi$ denotes the ratio of
the dark matter annihilation rate to the expansion rate immediately
after modulus decay, see Ref.~\cite{Lemoine:2006sc}. The above
provides reliable approximations of more complete formulas discussed
in Ref.~\cite{LMY09}. We quantify the amount of matter -- radiation
isocurvature fluctuation through $\delta_{\rm m\gamma}$:
\begin{equation}
  \delta_{\rm m\gamma}\,\equiv\, \frac{\zeta_{\rm
      m}-\zeta_\gamma}{\left(\zeta_{\rm
        m}+\zeta_{\gamma}\right)/2},
\label{eq:delta}
\end{equation}
where $\zeta_{\rm m}\,\equiv\,\Omega _{\rm b}/\Omega _{\rm
  m}\zeta_{\rm b} + \Omega _{\chi}/\Omega _{\rm m}\zeta_{\rm \chi}$
with $\Omega _{\rm m}\equiv \Omega _{\chi}+\Omega _{\rm b}$. Using the
results of Ref.~\cite{Gordon:2002gv}, one finds that various CMB data
imply $-0.12<\delta _{\rm m\gamma }<0.089$ at $95\%$ C.L. (with
$\Omega _{\chi}h^2\simeq 0.12$ and $\Omega _{\rm b}h^2\simeq 0.0225$).

Consider now a concrete case with a time independent potential
$V(\sigma)\,=\,m_\sigma^2\sigma^2/2$. Then one finds:
\begin{equation}
\frac{\Omega _{\sigma}^{<_{\rm d}}}{\Omega_\gamma^{<_{\rm d}}}
\,\simeq\, 6\times10^{10}\,\left(\frac{\sigma_{\rm inf}}{M_{\rm
    Pl}}\right)^2\left(\frac{m_\sigma}{100\,{\rm
    TeV}}\right)^{-3/2}\left(\frac{T_{\rm rh}}{10^9\,{\rm
    GeV}}\right)\ ,
\label{eq:rsig-quad}
\end{equation}
assuming that the modulus starts to oscillate in its potential (at
$H=m_\sigma$) before reheating, an assumption which is valid
throughout the present parameter space provided $T_{\rm rh}\,\lesssim
10^{10}\,$GeV~\cite{LMY09}.

\begin{figure}
  \centering
  \includegraphics[width=0.5\textwidth,clip=true]{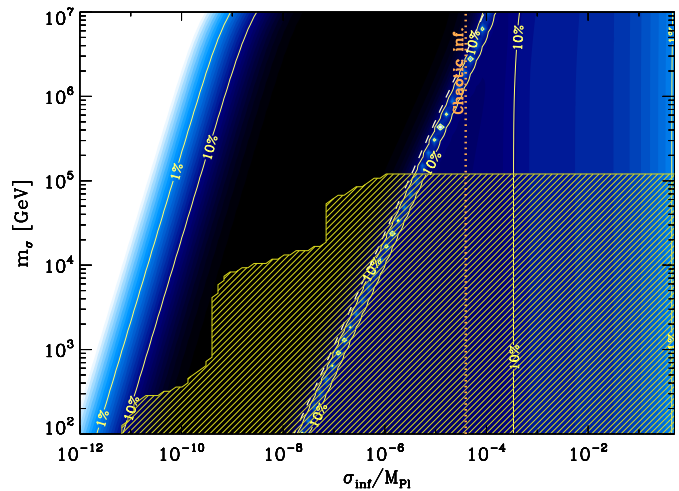}
  \caption[...]{Constraints in the ($\sigma_{\rm inf},\,m_\sigma$)
    plane. The yellow shaded region shows the constraints derived from
    the analysis of the effect of moduli decay on BBN. The blue
    contours give the value of $\delta_{\rm m \gamma}$ for the matter
    -- radiation isocurvature mode, constrained to be less than $\sim
    10\,$\% by CMB data. The white dashed line indicates the place
    where $\Omega_\sigma^{<_{\rm d}}=0.5$. These calculations assume
    $H_{\rm inf}=10^{13}\,$GeV and $T_{\rm rh}=10^9\,$GeV. The dotted
    orange line indicates the standard deviation of $\sigma$ on large
    scales due to stochastic motion during (chaotic $m^2\phi^2$)
    inflation.}
\label{fig:iso-quad-m2-tot}
\end{figure}

Figure~\ref{fig:iso-quad-m2-tot} presents the (blue colored) contours
of the $\delta_{\rm m\gamma}$ quantity calculated numerically through
the integration of the evolution equations for background and
perturbations as in Refs.~\cite{Lemoine:2006sc,Lemoine:2008qj},
assuming $T_{\rm rh}=10^9\,$GeV, $H_{\rm inf}=10^{13}\,$GeV; we set
$B_\chi=m_\chi/m_\sigma$ (\ie, one modulus particle produces one dark
matter particle at decay assuming instantaneous thermalization of the
dark matter, which is a good approximation as discussed
in~\cite{LMY09}) in regions of parameter space where the present day
dark matter density has the correct magnitude, and decrease $B_\chi$
accordingly in order to match this constraint
otherwise~\cite{LMY09}. The dashed yellow area is excluded by BBN; in
order to draw this region, we have used the results of
Ref.~\cite{Kawasaki:2004qu} for a hadronic branching ratio of
$10^{-3}$ and initial jet energy $1\,$TeV. One sees clearly that a
large portion of the parameter space is excluded by the generation of
isocurvature perturbations, well beyond the bounds of BBN. For $\sigma
_{\rm inf}\gta 4\times 10^{-6}(m_\sigma/100\,{\rm TeV})^{3/4}M_{\rm
  Pl}$ [with a numerical prefactor scaling as $(T_{\rm rh}/10^9\,{\rm
  GeV})^{-1/2}$], one finds $\Omega_{\sigma}^{<_{\rm d}}\simeq 1$
which leads to a strong baryon isocurvature mode but no dark
matter--radiation isocurvature perturbation. On the contrary, if
$\Omega_{\sigma}^{<_{\rm d}}$ is small and $\sigma_{\rm inf}\gta
10^{-10}(m_\sigma/100\,{\rm TeV})^{1/2}M_{\rm Pl}$ [with a numerical
prefactor scaling as $(T_{\rm rh}/10^9\,{\rm GeV})^{-1}$], there is no
baryon isocurvature mode but a large dark matter isocurvature mode. In
both regions, such large fractions should have been seen in current
CMB data. 

\par

Note that the typical displacement $\sigma_{\rm inf}$ of the inflaton
is bounded below by the standard deviation associated with the
stochastic motion of the modulus during inflation, whose magnitude is
quite large: $\langle\delta\sigma^2\rangle^{1/2}\,\sim\,H_{\rm
  inf}\sqrt{2}N_{_{\rm T}}/(2\pi)\,\gtrsim\,4\times10^{-5}M_{\rm Pl}$
for chaotic $m^2\phi^2$ inflation, with $N_{_{\rm T}}$ denotes the
total number of $e-$folds; the last inequality uses $H_{\rm
  inf}=10^{13}\,$GeV and $N_{_{\rm T}}\,\geq\,60$. This value is
indicated in dotted lines in Fig.~\ref{fig:iso-quad-m2-tot}. The above
estimate of $\langle\delta\sigma^2\rangle^{1/2}$ furthermore
represents a lower limit to $\sigma_{\rm inf}$, as the modulus may
have been initially displaced from its minimum. Actually, most of the
region above this bound is also excluded by the normalization of the
total curvature perturbation to CMB data (see the corresponding
discussion in ~\cite{LMY09}). \par

Let us note at this stage the implications for moduli cosmology and
model building. First of all, even if moduli are massive enough to
evade BBN constraints (\ie, $m_\sigma\,\gta 100\,$TeV), they are bound
to produce strong isocurvature fluctuations between baryons and
radiation (at high values of $\sigma_{\rm inf}$) or between dark
matter and radiation (at small values of $\sigma_{\rm inf}$).  
More generally, the present results (under the present assumptions)
forbid any large amount of entropy production by late time decaying
scalars, since this would precisely be accompanied by strong baryon
isocurvature fluctuations. This result bears strong implications for
the dilution of unwanted relics through moduli decay. The present
perturbative moduli problem requires to reduce quite significantly the
Hubble scale of inflation, as discussed at the end of this paper. This
would mean, of course, that tensor modes would become unaccessible to
the upcoming generation of CMB instruments.

\par

\begin{figure}
  \centering
  \includegraphics[width=0.5\textwidth,clip=true]{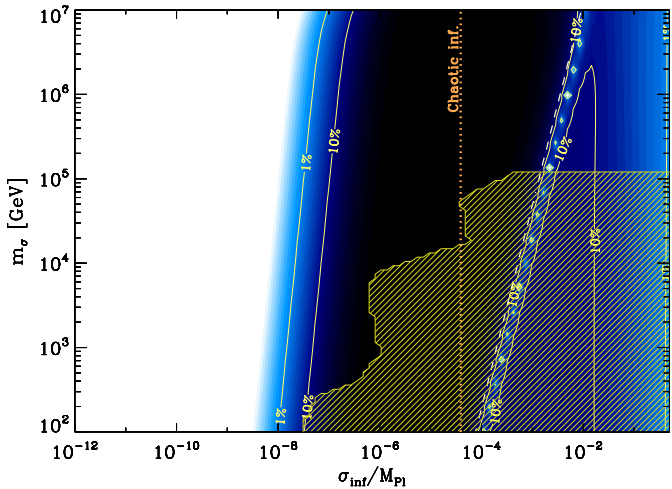}
  \caption[...]{Same as Fig.~\ref{fig:iso-quad-m2-tot}, for a modulus
    potential including supergravity corrections as in
    Eq.~(\ref{eq:effH}) with $c^2=0.5$.  } \label{fig:iso-H2-1}
\end{figure}

We now discuss a second prototypical case, in which the potential
receives supergravity corrections $V''\sim H^2$ associated with the
breaking of supersymmetry by the finite energy density in the early
Universe~\cite{Dine:1995uk}:
\begin{equation}
V(\sigma)\,\simeq\, \frac{1}{2}m_\sigma^2\sigma^2 +
\frac{1}{2}c^2H^2\sigma^2\ .\label{eq:effH}
\end{equation}
Such a potential is realized after inflation, for instance, in D-term
inflation if $\sigma$ corresponds to the supersymmetry breaking
Polonyi field, the field $\sigma$ remaining light during
inflation~\cite{LMY09}. The modulus abundance at the onset of
oscillations ($H=m_\sigma$) is smaller than in the previous case by a
factor $(m_\sigma/H_{\rm inf})^{2(\mu+1-\nu)}$, with $\mu=-1/2$,
$\nu^2=1/4 - 4c^2/9$ (we assume $c<3/4$ here). Consequently, the
constraints in the modulus parameter space shift to increasing values
of $\sigma_{\rm inf}$, as shown in Fig.~\ref{fig:iso-H2-1}.

Figure~\ref{fig:iso-H2-1} further establishes the power of constraints
obtained at the perturbative level, as it shows that the isocurvature
constraints still extend below the lower limit on $\sigma_{\rm inf}$
associated with stochastic motion of the modulus in its potential
during inflation. With respect to concrete moduli model building, even
if the minima of the modulus effective potentials at high and low
energy match one another, the isocurvature constraints close the
available parameter space for the present rather generic assumptions.
It has been suggested that the modulus may be kept close to the
minimum of the low energy potential when this latter is an enhanced
symmetry point in moduli space, thanks to the friction that results
from its coupling to other light degrees of
freedom~\cite{Dine:1995uk,Dine:2000ds,Kofman:2004yc}.  Whether this
can be realized during inflation lies beyond the scope of the present
work; it calls however for a detailed study of the stochastic motion
of the moduli in this context of realistic moduli potentials.

\par

\begin{figure}
  \centering
  \includegraphics[width=0.5\textwidth,clip=true]{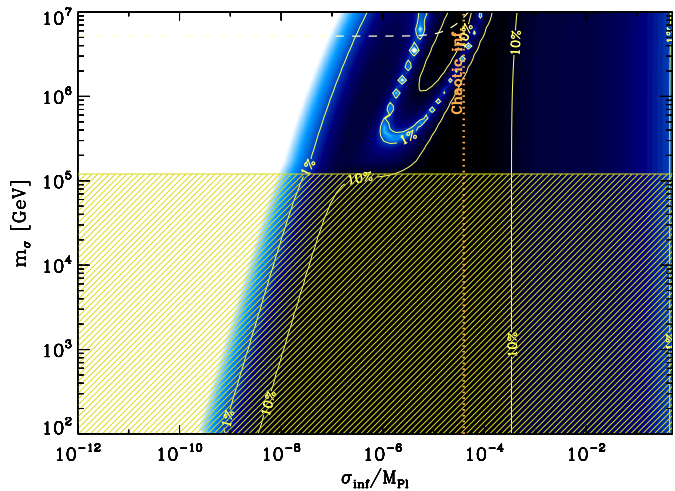}
  \caption[...]{Same as Fig.~\ref{fig:iso-H2-1}, accounting for
    inflaton produced moduli with $N_\sigma=10^{-3}$ and an inflaton
    mass $m_\phi=10^{13}\,$GeV.  } \label{fig:iso-H2ip-2}
\end{figure}

Finally, one must consider the possibility that the inflaton decays
partially into the modulus sector at reheating. This would attenuate
the amount of isocurvature fluctuations produced at the same value of
$\Omega_\sigma^{<_{\rm d}}$ as the inflaton produced moduli inherit
the same spectrum of perturbations than
radiation~\cite{Linde:2005yw}. Explicitly, $S_{\sigma\gamma}^{\rm
  (i)}$ is reduced by the factor $1/(1+\gamma')$, where $\gamma'$
represents the energy density ratio of inflaton produced moduli to
those initially present. Assuming that each inflaton of mass $m_\phi$
produces $N_\sigma$ moduli, one finds~\cite{LMY09}:
\begin{equation}
\gamma' \simeq  10^{-11}\,\Omega _{\rm \sigma, osci}^{-1}
\left(\frac{N_\sigma}{10^{-3}}\right)
\left(\frac{m_\sigma}{100\,{\rm TeV}}\right)
\left(\frac{m_\phi}{10^{13}\,{\rm GeV}}\right)^{-1}\ ,
\end{equation}
where $\Omega_{\sigma,\rm osci}\equiv \Omega_\sigma(H=m_\sigma)$, and
$\Omega_{\sigma ,\rm osci}=(\sigma_{\rm inf}/M_{\rm pl})^2/6$ for a
potential without supergravity corrections. If
$N_\sigma\,\sim\,g_{*}^{-1}\,\simeq\,10^{-3}$ and if the modulus
potential receives supergravity corrections after inflation with
$c^2=0.5$, one obtains the bounds depicted in
Fig.~\ref{fig:iso-H2ip-2}.  Compared to
Fig.~\ref{fig:iso-quad-m2-tot}, the BBN constraints now exclude all
moduli masses below $100\,$TeV. This is expected insofar as the amount
of moduli energy density produced through inflaton decay is already
sufficient to disrupt BBN. From this point of view, the production of
moduli through inflaton decay aggravates the moduli problem. The
contours depicting the amount of isocurvature fluctuations produced
are shifted toward higher values of $\sigma_{\rm inf}$, as a result of
the reduction of isocurvature fluctuations. However, they do not
vanish, and there remains a significant baryon-radiation isocurvature
mode in most of parameter space.

{\it Discussion.}  The decay of a modulus generically produces strong
isocurvature fluctuations between dark matter and radiation, or
between baryons and radiation. For a prototypical cosmological
inflationary scenario with $H_{\rm inf}=10^{13}\,$GeV, we find
that the moduli problem appears significantly worse at the
perturbative level. In a way, this opens a window on moduli
phenomenology for future high accuracy CMB experiments. As one of our
main results, we find that particle physics scenarios with heavy
moduli ($m_\sigma\gta \,100 \,$TeV) do not escape the above
constraints even though they evade BBN constraints. 

The detailed study reported in~\cite{LMY09} suggests that generically,
one needs to achieve $H_{\rm inf}\,\ll\,10^{13}\,$GeV in order to
solve the moduli problem, and this bears strong implications for the
possibility of detecting tensor modes in the CMB data with the
upcoming generation of instruments. In details, if $H_{\rm
  inf}\,\ll\,10^{13}\,$GeV, then the isocurvature mode
$S_{\sigma\gamma}^{\rm (i)}\,\propto\, H_{\rm inf}/\sigma_{\rm inf}$
becomes sufficiently small to evade detection at the \% level for
$\sigma_{\rm inf}\,\gg\, M_{\rm Pl}\, \left(H_{\rm inf}/10^{13}\,{\rm
  GeV}\right)$. Then a window opens at $m_\sigma\,\gtrsim\,100\,$TeV
(to evade BBN constraints) and $\sigma_{\rm inf}\,\sim\,M_{\rm
  Pl}$. As a concrete realization of this scenario, one can consider
the so-called KKLT potential for modulus
stabilization~\cite{Kachru:2003aw}, which yields
$m_\sigma\,\simeq\,700\,$TeV for their choice of numerical values, and
an initial displacement $\sigma_{\rm inf}\,\sim\,{\cal O}(0.1)\,M_{\rm
  Pl}$, with the KKLMMT inflationary model, which gives $H_{\rm
  inf}\,\sim\, 10^{9}\,$GeV~\cite{Kachru:2003sx}.

 Alternatively, if the modulus receives an Hubble effective mass
 $c_{\rm i}H$ during inflation, the isocurvature mode is erased during
 inflation~\cite{Yamaguchi:2005qm} hence the constraints obtained at
 the perturbative level vanish. However, the stochastic motion of the
 modulus in its potential is sufficiently large to disrupt BBN, if
 $c_{\rm i}$ is not significantly larger than unity or if $m_\sigma$
 is not larger than $100\,$TeV~\cite{LMY09}. Reducing the amplitude of
 this quantum noise then also requires lowering the scale of inflation
 well below $10^{13}\,$GeV.

A large amount of entropy production at low scales
$H\,\lesssim\,m_\sigma$ through heavy particle decay or late time
inflation could in principle help, by diluting the moduli abundance to
low levels~\cite{Randall:1994fr,Lyth:1995ka}. However, the present
constraints on entropy release by a modulus can be directly recast on
the decaying field or late time inflaton. Since the radiation produced in
the reheating inherits the fluctuations of the entropy producing
fluid, a strong baryon isocurvature fluctuation is produced unless
baryogenesis takes place subsequently to reheating,  
see~\cite{Lemoine:2008qj}. Given that the intermediate scale
$\sqrt{M_WM_{\rm Pl}}\,\simeq\, 10^{10}\,$GeV, this  brings in
additional non-trivial requirements.

\acknowledgments{This work was supported in part by CNRS-JSPS
  bilateral project of cooperative research and JSPS Grant-in-Aid for
  Scientific Research No.\ 19340054 (JY)}

\bibliography{references}

\begin{thebibliography}{23}
\expandafter\ifx\csname natexlab\endcsname\relax\def\natexlab#1{#1}\fi
\expandafter\ifx\csname bibnamefont\endcsname\relax
  \def\bibnamefont#1{#1}\fi
\expandafter\ifx\csname bibfnamefont\endcsname\relax
  \def\bibfnamefont#1{#1}\fi
\expandafter\ifx\csname citenamefont\endcsname\relax
  \def\citenamefont#1{#1}\fi
\expandafter\ifx\csname url\endcsname\relax
  \def\url#1{\texttt{#1}}\fi
\expandafter\ifx\csname urlprefix\endcsname\relax\def\urlprefix{URL }\fi
\providecommand{\bibinfo}[2]{#2}
\providecommand{\eprint}[2][]{\url{#2}}

\bibitem[{\citenamefont{Coughlan et~al.}(1983)\citenamefont{Coughlan, Fischler,
  Kolb, Raby, and Ross}}]{Coughlan:1983ci}
\bibinfo{author}{\bibfnamefont{G.~D.} \bibnamefont{Coughlan}},
  \bibinfo{author}{\bibfnamefont{W.}~\bibnamefont{Fischler}},
  \bibinfo{author}{\bibfnamefont{E.~W.} \bibnamefont{Kolb}},
  \bibinfo{author}{\bibfnamefont{S.}~\bibnamefont{Raby}}, \bibnamefont{and}
  \bibinfo{author}{\bibfnamefont{G.~G.} \bibnamefont{Ross}},
  \bibinfo{journal}{Phys. Lett.} \textbf{\bibinfo{volume}{B131}},
  \bibinfo{pages}{59} (\bibinfo{year}{1983}).

\bibitem[{\citenamefont{Goncharov et~al.}(1984)\citenamefont{Goncharov, Linde,
  and Vysotsky}}]{Goncharov:1984qm}
\bibinfo{author}{\bibfnamefont{A.~S.} \bibnamefont{Goncharov}},
  \bibinfo{author}{\bibfnamefont{A.~D.} \bibnamefont{Linde}}, \bibnamefont{and}
  \bibinfo{author}{\bibfnamefont{M.~I.} \bibnamefont{Vysotsky}},
  \bibinfo{journal}{Phys. Lett.} \textbf{\bibinfo{volume}{B147}},
  \bibinfo{pages}{279} (\bibinfo{year}{1984}).

\bibitem[{\citenamefont{Kawasaki et~al.}(2005)\citenamefont{Kawasaki, Kohri,
  and Moroi}}]{Kawasaki:2004qu}
\bibinfo{author}{\bibfnamefont{M.}~\bibnamefont{Kawasaki}},
  \bibinfo{author}{\bibfnamefont{K.}~\bibnamefont{Kohri}}, \bibnamefont{and}
  \bibinfo{author}{\bibfnamefont{T.}~\bibnamefont{Moroi}},
  \bibinfo{journal}{Phys. Rev.} \textbf{\bibinfo{volume}{D71}},
  \bibinfo{pages}{083502} (\bibinfo{year}{2005}), \eprint{astro-ph/0408426}.

\bibitem[{\citenamefont{Randall and Thomas}(1995)}]{Randall:1994fr}
\bibinfo{author}{\bibfnamefont{L.}~\bibnamefont{Randall}} \bibnamefont{and}
  \bibinfo{author}{\bibfnamefont{S.~D.} \bibnamefont{Thomas}},
  \bibinfo{journal}{Nucl. Phys.} \textbf{\bibinfo{volume}{B449}},
  \bibinfo{pages}{229} (\bibinfo{year}{1995}), \eprint{hep-ph/9407248}.

\bibitem[{\citenamefont{Lyth and Stewart}(1996)}]{Lyth:1995ka}
\bibinfo{author}{\bibfnamefont{D.~H.} \bibnamefont{Lyth}} \bibnamefont{and}
  \bibinfo{author}{\bibfnamefont{E.~D.} \bibnamefont{Stewart}},
  \bibinfo{journal}{Phys. Rev.} \textbf{\bibinfo{volume}{D53}},
  \bibinfo{pages}{1784} (\bibinfo{year}{1996}), \eprint{hep-ph/9510204}.

\bibitem[{\citenamefont{Dine et~al.}(1995)\citenamefont{Dine, Randall, and
  Thomas}}]{Dine:1995uk}
\bibinfo{author}{\bibfnamefont{M.}~\bibnamefont{Dine}},
  \bibinfo{author}{\bibfnamefont{L.}~\bibnamefont{Randall}}, \bibnamefont{and}
  \bibinfo{author}{\bibfnamefont{S.~D.} \bibnamefont{Thomas}},
  \bibinfo{journal}{Phys. Rev. Lett.} \textbf{\bibinfo{volume}{75}},
  \bibinfo{pages}{398} (\bibinfo{year}{1995}), \eprint{hep-ph/9503303}.

\bibitem[{\citenamefont{Dine}(2000)}]{Dine:2000ds}
\bibinfo{author}{\bibfnamefont{M.}~\bibnamefont{Dine}}, \bibinfo{journal}{Phys.
  Lett.} \textbf{\bibinfo{volume}{B482}}, \bibinfo{pages}{213}
  (\bibinfo{year}{2000}), \eprint{hep-th/0002047}.

\bibitem[{\citenamefont{Ellis et~al.}(1986)\citenamefont{Ellis, Nanopoulos, and
  Quiros}}]{Ellis:1986zt}
\bibinfo{author}{\bibfnamefont{J.~R.} \bibnamefont{Ellis}},
  \bibinfo{author}{\bibfnamefont{D.~V.} \bibnamefont{Nanopoulos}},
  \bibnamefont{and} \bibinfo{author}{\bibfnamefont{M.}~\bibnamefont{Quiros}},
  \bibinfo{journal}{Phys. Lett.} \textbf{\bibinfo{volume}{B174}},
  \bibinfo{pages}{176} (\bibinfo{year}{1986}).

\bibitem[{\citenamefont{Kawasaki et~al.}(1996)\citenamefont{Kawasaki, Moroi,
  and Yanagida}}]{Kawasaki:1995cy}
\bibinfo{author}{\bibfnamefont{M.}~\bibnamefont{Kawasaki}},
  \bibinfo{author}{\bibfnamefont{T.}~\bibnamefont{Moroi}}, \bibnamefont{and}
  \bibinfo{author}{\bibfnamefont{T.}~\bibnamefont{Yanagida}},
  \bibinfo{journal}{Phys. Lett.} \textbf{\bibinfo{volume}{B370}},
  \bibinfo{pages}{52} (\bibinfo{year}{1996}), \eprint{hep-ph/9509399}.

\bibitem[{\citenamefont{Moroi and Randall}(2000)}]{Moroi:1999zb}
\bibinfo{author}{\bibfnamefont{T.}~\bibnamefont{Moroi}} \bibnamefont{and}
  \bibinfo{author}{\bibfnamefont{L.}~\bibnamefont{Randall}},
  \bibinfo{journal}{Nucl. Phys.} \textbf{\bibinfo{volume}{B570}},
  \bibinfo{pages}{455} (\bibinfo{year}{2000}), \eprint{hep-ph/9906527}.

\bibitem[{\citenamefont{Acharya et~al.}(2008)}]{Acharya:2008bk}
\bibinfo{author}{\bibfnamefont{B.~S.} \bibnamefont{Acharya}}
  \bibnamefont{et~al.}, \bibinfo{journal}{JHEP} \textbf{\bibinfo{volume}{06}},
  \bibinfo{pages}{064} (\bibinfo{year}{2008}), \eprint{0804.0863}.

\bibitem[{\citenamefont{Moroi and Takahashi}(2001)}]{Moroi:2001ct}
\bibinfo{author}{\bibfnamefont{T.}~\bibnamefont{Moroi}} \bibnamefont{and}
  \bibinfo{author}{\bibfnamefont{T.}~\bibnamefont{Takahashi}},
  \bibinfo{journal}{Phys. Lett.} \textbf{\bibinfo{volume}{B522}},
  \bibinfo{pages}{215} (\bibinfo{year}{2001}), \eprint{hep-ph/0110096}.

\bibitem[{\citenamefont{Boyle et~al.}(2006)\citenamefont{Boyle, Steinhardt, and
  Turok}}]{Boyle:2005ug}
\bibinfo{author}{\bibfnamefont{L.~A.} \bibnamefont{Boyle}},
  \bibinfo{author}{\bibfnamefont{P.~J.} \bibnamefont{Steinhardt}},
  \bibnamefont{and} \bibinfo{author}{\bibfnamefont{N.}~\bibnamefont{Turok}},
  \bibinfo{journal}{Phys. Rev. Lett.} \textbf{\bibinfo{volume}{96}},
  \bibinfo{pages}{111301} (\bibinfo{year}{2006}), \eprint{astro-ph/0507455}.

\bibitem[{\citenamefont{Lemoine et~al.}(2009)\citenamefont{Lemoine, Martin, and
  Yokoyama}}]{LMY09}
\bibinfo{author}{\bibfnamefont{M.}~\bibnamefont{Lemoine}},
  \bibinfo{author}{\bibfnamefont{J.}~\bibnamefont{Martin}}, \bibnamefont{and}
  \bibinfo{author}{\bibfnamefont{J.}~\bibnamefont{Yokoyama}}
  (\bibinfo{year}{2009}).

\bibitem[{\citenamefont{Langlois and Vernizzi}(2004)}]{Langlois:2004nn}
\bibinfo{author}{\bibfnamefont{D.}~\bibnamefont{Langlois}} \bibnamefont{and}
  \bibinfo{author}{\bibfnamefont{F.}~\bibnamefont{Vernizzi}},
  \bibinfo{journal}{Phys. Rev.} \textbf{\bibinfo{volume}{D70}},
  \bibinfo{pages}{063522} (\bibinfo{year}{2004}), \eprint{astro-ph/0403258}.

\bibitem[{\citenamefont{Lemoine and Martin}(2007)}]{Lemoine:2006sc}
\bibinfo{author}{\bibfnamefont{M.}~\bibnamefont{Lemoine}} \bibnamefont{and}
  \bibinfo{author}{\bibfnamefont{J.}~\bibnamefont{Martin}},
  \bibinfo{journal}{Phys. Rev.} \textbf{\bibinfo{volume}{D75}},
  \bibinfo{pages}{063504} (\bibinfo{year}{2007}), \eprint{astro-ph/0611948}.

\bibitem[{\citenamefont{Gordon and Lewis}(2003)}]{Gordon:2002gv}
\bibinfo{author}{\bibfnamefont{C.}~\bibnamefont{Gordon}} \bibnamefont{and}
  \bibinfo{author}{\bibfnamefont{A.}~\bibnamefont{Lewis}},
  \bibinfo{journal}{Phys. Rev.} \textbf{\bibinfo{volume}{D67}},
  \bibinfo{pages}{123513} (\bibinfo{year}{2003}), \eprint{astro-ph/0212248}.

\bibitem[{\citenamefont{Lemoine et~al.}(2008)\citenamefont{Lemoine, Martin, and
  Petit}}]{Lemoine:2008qj}
\bibinfo{author}{\bibfnamefont{M.}~\bibnamefont{Lemoine}},
  \bibinfo{author}{\bibfnamefont{J.}~\bibnamefont{Martin}}, \bibnamefont{and}
  \bibinfo{author}{\bibfnamefont{G.}~\bibnamefont{Petit}},
  \bibinfo{journal}{Phys. Rev.} \textbf{\bibinfo{volume}{D78}},
  \bibinfo{pages}{063516} (\bibinfo{year}{2008}), \eprint{0802.1601}.

\bibitem[{\citenamefont{Kofman et~al.}(2004)}]{Kofman:2004yc}
\bibinfo{author}{\bibfnamefont{L.}~\bibnamefont{Kofman}} \bibnamefont{et~al.},
  \bibinfo{journal}{JHEP} \textbf{\bibinfo{volume}{05}}, \bibinfo{pages}{030}
  (\bibinfo{year}{2004}), \eprint{hep-th/0403001}.

\bibitem[{\citenamefont{Linde and Mukhanov}(2006)}]{Linde:2005yw}
\bibinfo{author}{\bibfnamefont{A.}~\bibnamefont{Linde}} \bibnamefont{and}
  \bibinfo{author}{\bibfnamefont{V.}~\bibnamefont{Mukhanov}},
  \bibinfo{journal}{JCAP} \textbf{\bibinfo{volume}{0604}}, \bibinfo{pages}{009}
  (\bibinfo{year}{2006}), \eprint{astro-ph/0511736}.

\bibitem[{\citenamefont{Kachru et~al.}(2003{\natexlab{a}})\citenamefont{Kachru,
  Kallosh, Linde, and Trivedi}}]{Kachru:2003aw}
\bibinfo{author}{\bibfnamefont{S.}~\bibnamefont{Kachru}},
  \bibinfo{author}{\bibfnamefont{R.}~\bibnamefont{Kallosh}},
  \bibinfo{author}{\bibfnamefont{A.}~\bibnamefont{Linde}}, \bibnamefont{and}
  \bibinfo{author}{\bibfnamefont{S.~P.} \bibnamefont{Trivedi}},
  \bibinfo{journal}{Phys. Rev.} \textbf{\bibinfo{volume}{D68}},
  \bibinfo{pages}{046005} (\bibinfo{year}{2003}{\natexlab{a}}),
  \eprint{hep-th/0301240}.

\bibitem[{\citenamefont{Kachru et~al.}(2003{\natexlab{b}})}]{Kachru:2003sx}
\bibinfo{author}{\bibfnamefont{S.}~\bibnamefont{Kachru}} \bibnamefont{et~al.},
  \bibinfo{journal}{JCAP} \textbf{\bibinfo{volume}{0310}}, \bibinfo{pages}{013}
  (\bibinfo{year}{2003}{\natexlab{b}}), \eprint{hep-th/0308055}.

\bibitem[{\citenamefont{Yamaguchi and Yokoyama}(2006)}]{Yamaguchi:2005qm}
\bibinfo{author}{\bibfnamefont{M.}~\bibnamefont{Yamaguchi}} \bibnamefont{and}
  \bibinfo{author}{\bibfnamefont{J.}~\bibnamefont{Yokoyama}},
  \bibinfo{journal}{Phys. Rev.} \textbf{\bibinfo{volume}{D74}},
  \bibinfo{pages}{043523} (\bibinfo{year}{2006}), \eprint{hep-ph/0512318}.

\end{thebibliography}

\end{document}